\documentclass[aps,pre,twocolumn,showpacs,floatfix]{revtex4-1}
\usepackage{amsmath,amsfonts,bm,color,amssymb}
\usepackage{graphicx,epsfig,overpic,graphics,epstopdf}

\usepackage{epstopdf}

\begin{document}
\title{Experimental Studies of Two-dimensional Laminar Jet Flows in Freely Suspended Liquid Crystal Films}

\date{\today}
\author{Kyle R.~Ferguson\textsuperscript{1, 2}}
\author{Evan Dutch\textsuperscript{1, 2}}
\author{Zhiyuan Qi\textsuperscript{1, 2}}
\author{Adam Green\textsuperscript{1, 2}}
\author{Carlos Alas\textsuperscript{3}}
\author{Corrina Briggs\textsuperscript{1, 2}}
\author{Cheol Soo Park\textsuperscript{1, 2}}
\author{Matthew A.~Glaser\textsuperscript{1, 2}}
\author{Joseph E.~Maclennan\textsuperscript{1, 2}}
\author{Noel A.~Clark\textsuperscript{1, 2}}
\affiliation{Department of Physics\textsuperscript{1}and Soft Materials Research Center\textsuperscript{2}, University of Colorado, Boulder, Colorado, 80309, USA}
\affiliation{\textsuperscript{3}Department of Physics, California Polytechnic State University, San Luis Obispo, California, 93407, USA}
\date{\today}

\begin{abstract}

Two dimensional (2D) laminar jet-----a stream of fluid that projected into a surrounding medium with the flow confined in 2D-----has both theoretical and experimental significance. We carried out 2D laminar jet experiments in freely suspended liquid crystal films (FSLCFs) of nanometers thick and
centimeters in size, in which individual molecules are confined to single layers thus enable film flows with two degrees of freedom. The experimental observations are found in good agreement with the classic 2D laminar jet theory of ideal cases that assume no external coupling effects, even in fact there exist strong coupling force from the ambient air. We further investigated this air coupling effect in computer simulations, with the results indicated air has little influence on the velocity maps of flow near the nozzle. This astonishing results could be intuitively understood by considering 2D incompressibility of the films. This experiment, together with a series of our previous experiments, show for a wide range of Reynolds number, FSLCFs are excellent testing beds for  2D hydrodynamics.
\end{abstract}


\maketitle

\section{Introduction}
Two dimensional hydrodynamics has attracted the attention of both theorists and experimentalists for more than a century~\cite{Page_1913,Kuo_1949,Weiss_1991,monnier2016inverse}. One reason behind this is the simplicity of two dimensional flows, which can give good insights into a fully understanding of their counterparts in three dimensions, while capable of maintaining a relatively easier analytical or numerical solution as a comparison~\cite{anderson1995computational}. Another reason is that the two dimensional flows have their own importance and applicability in both research~\cite{weiss1991dynamics} and engineering~\cite{sedov1980two,park2014developing}. 

The experimental verification of two dimensional hydrodynamics, however, was heavily restricted by the basic fact that any fluids in nature are intrinsically three dimensions~\cite{lamb1932hydrodynamics}. Fluids are composed by large quantity of atoms and molecules which have size and volume. To overcome this difficulty, two dimensional hydrodynamic experiments were carried out in water with wire falling sideways in viscous fluids~\cite{white1946drag}; this large length-to-diameter ratio enables flow at positions along the wire axis maintaining some similarities, making it possible to mimic two dimensional flows using three dimensional fluids. Some early hydrodynamic experiment on a Karman Vortex Street, such as by Fage and Johansen~\cite{fage1928xlii}, were carried out in a similar approach where in a water channel a long cylindrical shape with axis  normal to the direction of motion, the flow will be the same in all planes normal  to the axis, and may be conceived as proceeding in two dimension~\cite{glauert1928characteristics}. The experimental study of the two dimensional inverse energy cascade in a square box by Sommeria~\cite{Sommeria_1986}, however, were implemented by driving a horizontal layer of mercury and by suppressing the three-dimensional perturbation by means of a uniform magnetic field. Those experiments gave out great insights into the understanding of two dimensional hydrodynamics. The mimicking of two dimensional flow by confining fluids moving evenly at different cross sections has many disadvantages. The experimental setups are typically heavy, complex and large in volume, which require large quantity of fluids. Also, due to the frictional force between the container boundaries and the fluids, this method can lead to large errors for some experiments. What's even worse, high speed flows can lead to unexpected turbulence which can easily destroy the evenly laminar flow behavior among different layers ~\cite{dashwu1997}.

\begin{figure}
\centering
\includegraphics[keepaspectratio=true,width=0.5\textwidth]{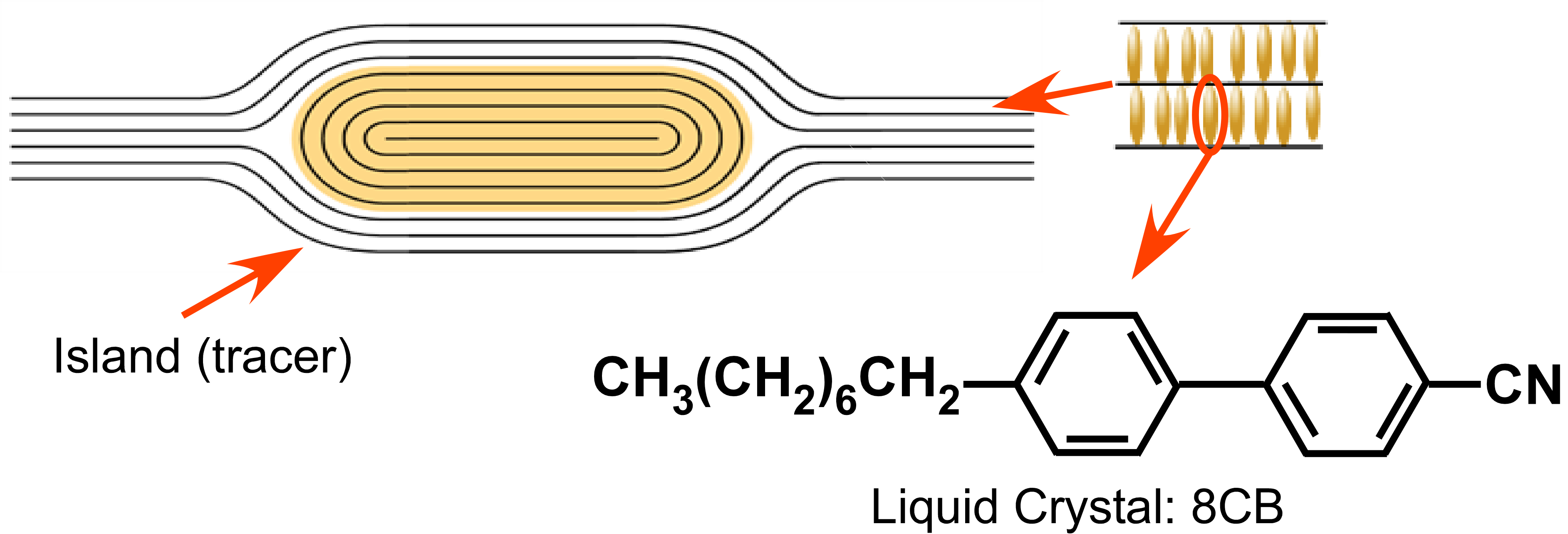}
\caption{A side view of an LC film with an island and the molecular structure of 8CB.}
\label{island}
\end{figure}

To overcome the difficulties as described above, liquid films such as the soap films, which have a very high degree of two-dimensionality~\cite{mysels1959frankel} were used as an experimental testing beds for two dimensional hydrodynamics~\cite{couder1989hydrodynamics,gharib1989liquid}. Many experiments such as on two dimensional velocity profile and boundary layers~\cite{rutgers1996two}, 2D Karman vortex street~\cite{afenchenko1998generation} and turbulence~\cite{vorobieff1999soap}, et al. were successfully implemented. However, the facial compressible nature of the soap films, causes the thickness of soap films tend to change when under frictional or gravitational force~\cite{gharib1989liquid}. Also, water evaporation causes unexpected flows, which can contaminate the accuracy of measurements ~\cite{kellay2017}.  

There is a long history of studying thin lipid membranes under the lens of 2D hydrodynamics~\cite{saffman_brownian_1975,hughes_translational_1981,petrov_translational_2008,petrov_translational_2012}, and results have shown that the field accurately applies to these materials. However, biological membranes are typically crowded, being populated by a high density of mutually hydrodynamically interacting proteins or protein assemblies, which prevents an accurate and precise understanding of hydrodynamic interaction within the membranes.  Recent studies~\cite{carbajal-tinoco_asymmetry_2007,prasad_flow_2009,nguyen_crossover_2010,eremin_two-dimensional_2011,Qi_mutual_2014} have led us to more closely examine the idea that smectic freely suspended liquid crystal films (FSLCFs) of nanometers thick and centimeters in size can act as ideal 2D fluids.  By the very nature of the films, individual molecules are confined to a single layer (Fig.~\ref{island}), and thus the flows within the film have only two degrees of freedom.  Additionally, the large surface-area-to-thickness ratio means that any flow between layers is negligible compared to the lateral flow. Finally, the low evaporation rate for LC films under room temperature and air pressure, leads to negligible unexpected flow in comparison with the case of soap films~\cite{qi2016experimental}.  

Our previous studies of FSLCFs focus on low Reynolds number 2D hydrodynamic behaviors, such as the crossover experiments from 2D to 3D transition~\cite{nguyen_crossover_2010}, 2D hydrodynamic interaction experiments between inclusion pairs~\cite{Qi_mutual_2014,kuriabova2016hydrodynamic}, experimental realization of 2D Newtonian fluids~\cite{qi2016experimental}, and active microrheological studies of smectic membranes~\cite{qi2017active}, little effort had been devoted to the verification of whether FSLCFs can still be regarded as ideal experimental testing beds for high Reynolds number cases. In this paper, we performed an experiment of two dimensional nozzle using FSLCFs, in which the fluid is injected at high momentum into a stationary reservoir of the same density. This experiment give us good opportunity for a precise measurement of 2D laminar jet flow, and allows us a concrete and quantitative understanding of the capabilities and limitations of using smectic LC films as a model for 2D fluids.

\section{Experiment}

\begin{figure}
\centering
\includegraphics[keepaspectratio=true,width=0.5\textwidth]{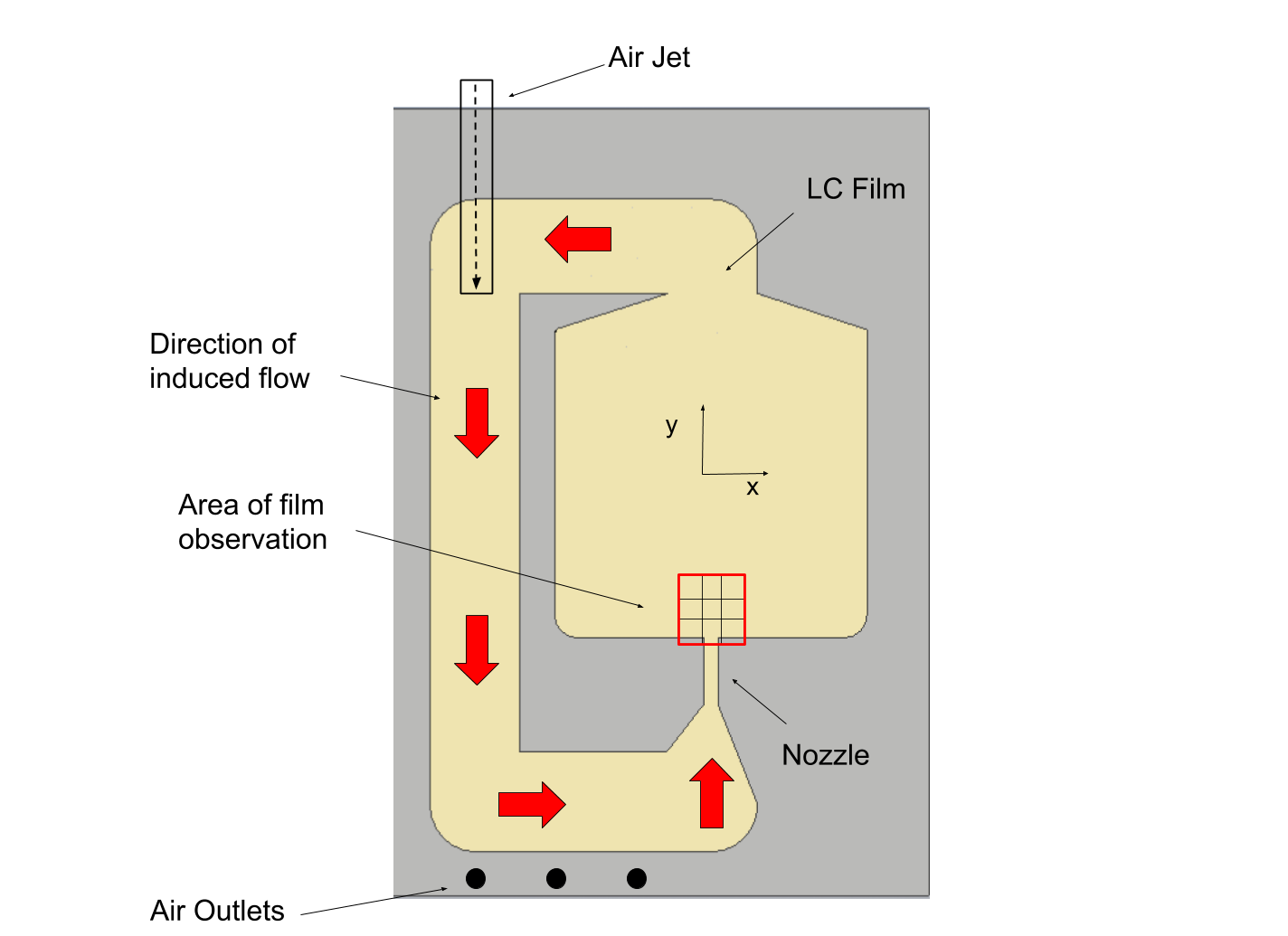}
\caption{ Schematic of the film holder.  Flow is excited by blowing air in the direction given by the black dotted arrow onto the film through a thin needle.  Video is captured in each of the nine black sub-regions within the larger red region, and the resulting flow fields are stitched together and analyzed.  The material used is $8$CB, which is in the Smectic A phase at room temperature.  The nozzle is $1~\mathrm{mm}$ wide by $4.25~\mathrm{mm}$, and the reservoir is $22~\mathrm{mm}$ wide by $21~\mathrm{mm}$ long.}
\label{channel_schematic}
\end{figure}

In order to implement laminar jet experiment using FSLCFs, we machined a film holder out of aluminum block with its schematic shown in Fig.~\ref{channel_schematic}. Note the yellow color stands for the channel area where the film sits, while the grew color are area of non-machined aluminum. By spreading a small amount of the liquid crystal material across this film holder in a glass cover slip, we manage to make a complex shaped film (yellow area) that basically composed of a reservoir, a 'C' shaped channel, and a nozzle. Air is blown at a controlled speed through a needle onto the film channel in a direction represented by the red line.  Due to the incompressibility of the film, this generates the long-range flow that follows the red arrows.  As the fluid travels around the track, it is funneled through the thin nozzle and into the large reservoir.  The reservoir was designed large enough such that any vorticity in the flow stays away from the nozzle thus has a negligible affect on flow field in the area observed, allowing us to treat the reservoir as infinite and the flow as irrotational in our theoretical analysis.  

The liquid crystal material used in our experiment is $8$CB (4\ensuremath{'}-n-octyl-4\ensuremath{'}-cyanobiphenyl, Sigma-Aldrich), which is in the fluid smectic~A phase at room temperature. The film thickness $h$, an integral number $N$ of smectic layers (typically $2 \le N \le 6$, each smectic layer is $3.17\,{\rm nm}$ thick), is determined precisely by comparing the reflectivity of the film with black glass. Immediately after a film is drawn, one typically observes many thicker islands floating on the film (Fig.~\ref{island}). Those islands were quickly broken into smaller ones under the air jet. The velocities of these islands are tracked in order to map out the flow field for the interested region.

The islands were then observed using reflected light video microscopy and taped using a high speed camera (Phantom~V$12.1$, Vision Research, Inc.) at a video frame rate of $5000$~fps. The obtained videos were then decomposed into sequential images and the flow fields were extracted from the tracer islands using velocimetric method~\cite{trackpy}. Due to the pixel limitation of the high speed camera, we can only capture a fraction of the nozzle flow region we intend to study. In order to capture the entire region of interest, nine videos were captured in a $3 \times 3$ square (shown in Fig.~\ref{channel_schematic}), each sub-region separated by a distance of $1.97~\mathrm{mm}$.  Each of the $9$ videos are taken at the same frame rate.  A computer controlled XY-stage was used to shift between different sub-regions. The flow field in each region was mapped by utilizing the particle tracking package Trackpy ~\cite{trackpy}. The flow fields of the $9$ regions were then stitched together into a single one.

\begin{figure}
\centering
\includegraphics[keepaspectratio=true,width=0.5\textwidth]{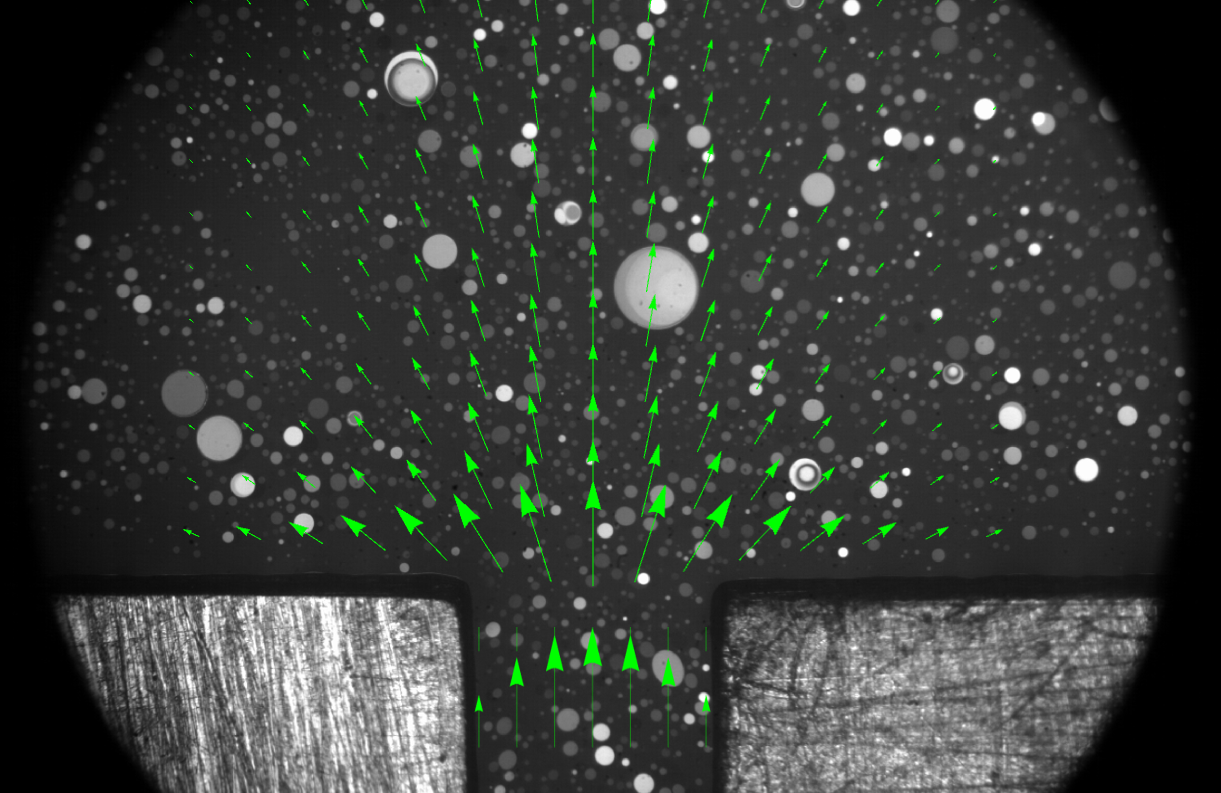}
\caption{An image of one of the nine regions. The bright discs are the islands whose velocities are used to map out the flow field of each region. The arrows represent the velocity vector feild.}
\label{region2}
\end{figure}

\section{Theory}

Although we are dealing with a viscous membrane, the dynamic viscosity of $8$CB is extremely small ($0.052 ~\mathrm{Pa} *\mathrm{sec}$); thus, in our theoretical treatment, we elect to ignore viscous effects.  Because we also make the assumption that the flow is irrotational, we may treat the system using the idea of $2$D complex potential flows.  Similar to the idea of a scalar potential in electrostatics, irrotational, inviscid velocity fields can be represented by a complex scalar function, $F(z)=\phi(x,y)+i\psi(x,y)$, where $z=x+iy=re^{i\theta}$ and $\phi$ and $\psi$ are both real functions.  $\phi$ is known as the velocity potential and $\psi$ as the stream function.  The flow will follow lines where $\psi=const.$  Velocity components may then be found by $v_x = \frac{\partial \phi}{\partial x} = \frac{\partial \psi}{\partial y}$ and $v_y = \frac{\partial \phi}{\partial y} = -\frac{\partial \psi}{\partial x}$, or by differentiating the complex potential to get the complex velocity, $G(z) = \frac{dF}{dz} = v_x - iv_y$.

When combined with the idea of conformal transformations $w=f(z)=u+iv$, which map points in the $z$-plane to new coordinates in the $w$-plane, this method becomes very powerful, as it can be used to describe flows of complex geometries.  We use a Schwarz-Christoffel transformation, which is actually defined by $\frac{dw}{dz}$,

\begin{equation}
\frac{dw}{dz} = \frac{2}{\pi}\frac{1}{(z-l)^{-1/2}(z-0)^{1}(z+l)^{-1/2}},
\end{equation}

\noindent or, integrating with respect to $z$,

\begin{equation}
w = \frac{2}{\pi}\bigg[i(l^2-z^2)^{1/2} + l\arctan{\frac{l}{i(l^2-z^2)^{1/2}}} \bigg].
\end{equation}

\begin{figure}
\centering
\includegraphics[keepaspectratio=true,width=0.5\textwidth]{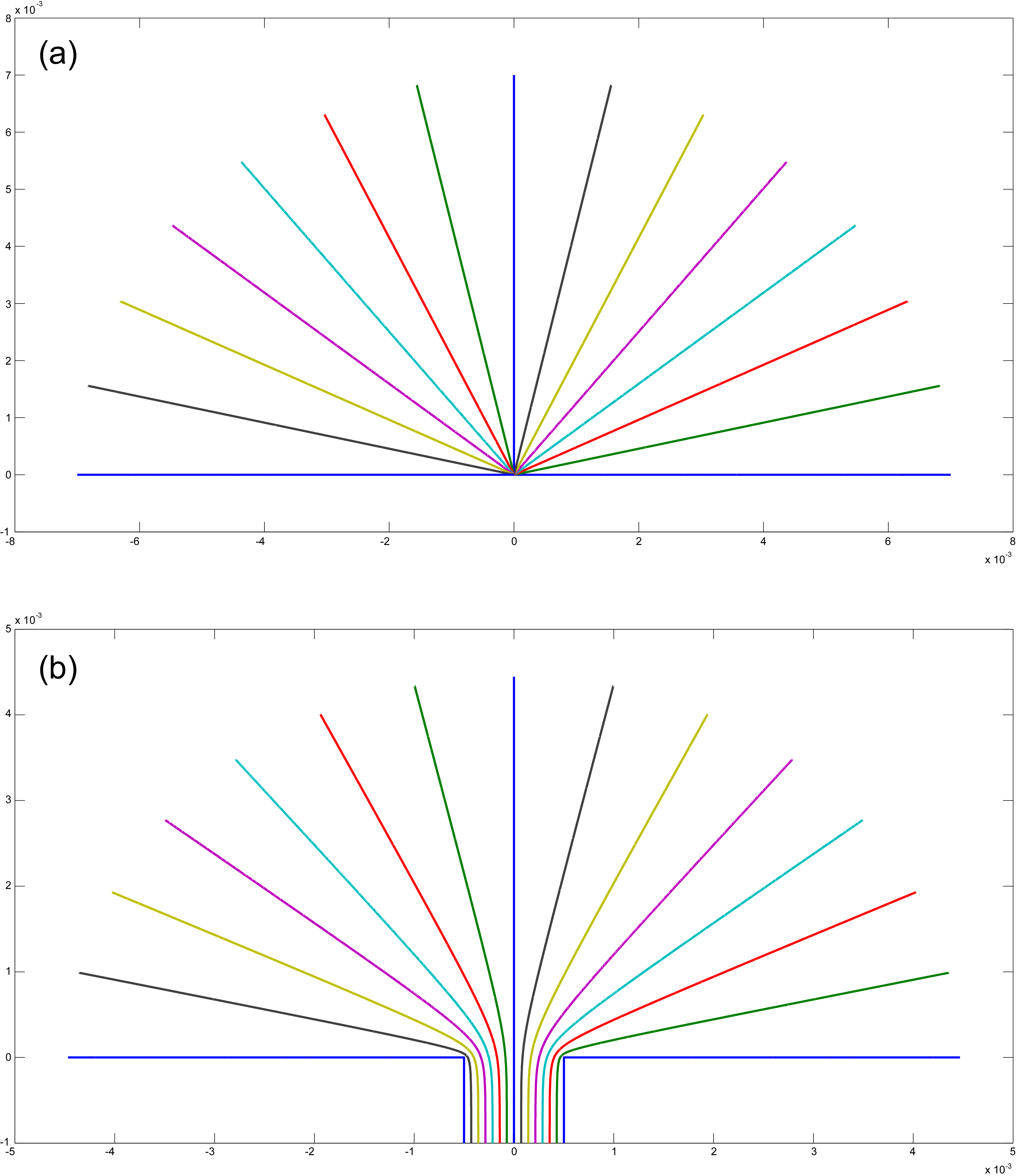}
\caption{(a) The streamlines for a source of flow given by the complex potential $F(z) = \frac{m}{2\pi}\ln{z}$.  (b) Mapping the streamlines through the Schwarz-Christoffel transformation $w = \frac{2}{\pi}\big[i(l^2-z^2)^{1/2} + l\arctan{\frac{l}{i(l^2-z^2)^{1/2}}} \big]$ yields flow through a nozzle into a semi-infinite reservoir.  Note that $2l$ is the nozzle width.}
\label{streamlines}
\end{figure}

\noindent This transformation maps the upper half of the $z$-plane, $-\infty < x < \infty, y > 0$, to the upper half of the $w$-plane, $-\infty < u < \infty, v > 0$, plus the strip representing the nozzle, $-l < u < l, v < 0$.
\begin{figure}
\centering
\includegraphics[keepaspectratio=true,width=0.5\textwidth]{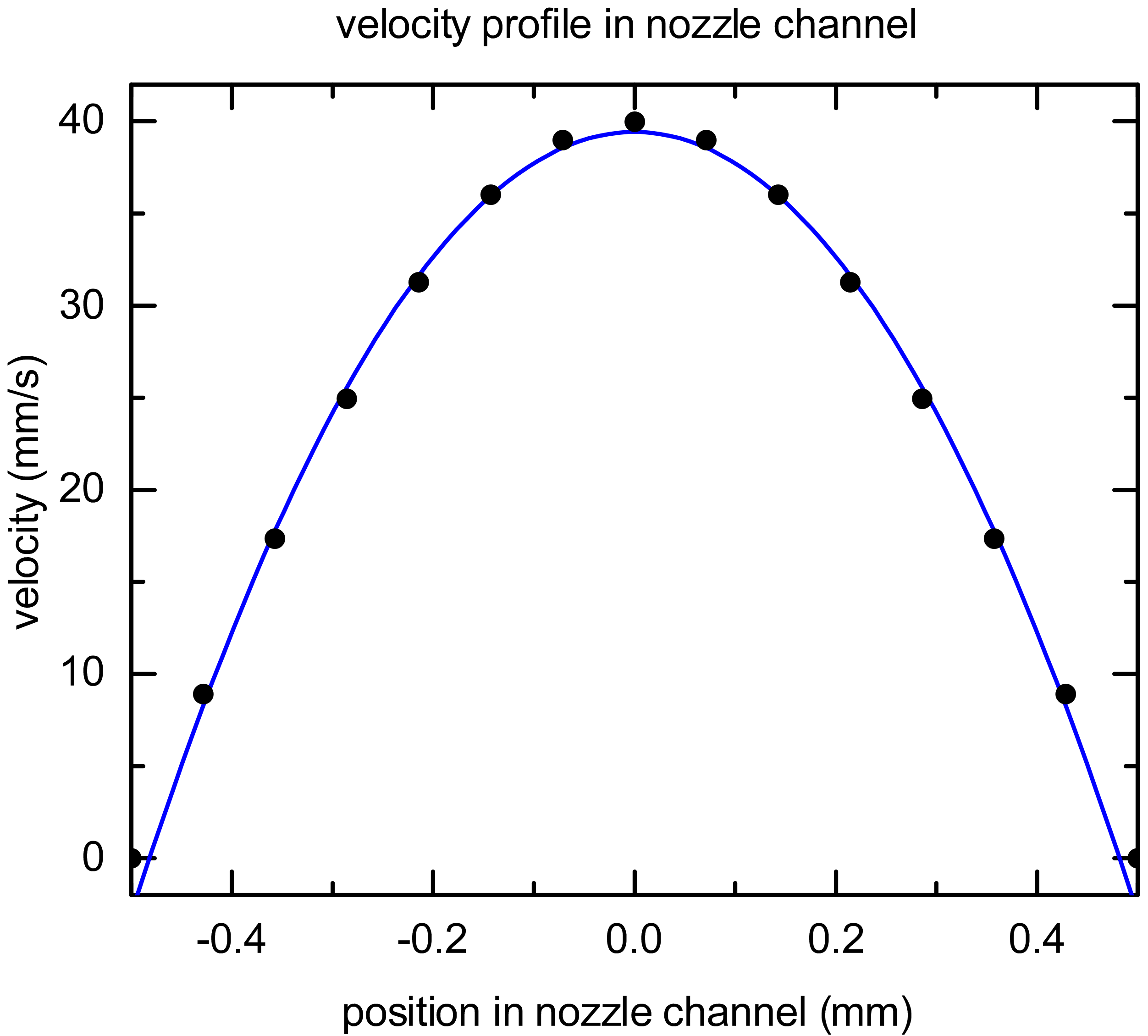}
\caption{Velocity profile in the nozzle channel when using a source that is non-uniform in strength.  The black points represent model-predicted velocities, and the blue line represents a parabolic best-fit.  The velocity profile is nearly quadratic in behavior, as previous experiments have shown it should be.}
\label{channel_profile}
\end{figure}
We examine a potential $F(z) = \frac{m}{2\pi}\ln{z}$.  This represents a source of strength $m$ sitting at the origin in the $z$-plane.  It has streamlines pointing straight outward ($m>0$) or inward ($m<0$) radially, as shown in Fig.~\ref{streamlines}~(a).  If we map the streamlines of this source through the transformation shown above, we see the streamlines of flow through a nozzle of width $2l$, shown in Fig.~\ref{streamlines}~(b).

Previous experimental work~\cite{qi2012} has demonstrated that the velocity profile in a straight channel is nearly quadratic in nature, with the maximum velocity being achieved in the center and no-slip boundary conditions on the side.  However, transforming a source of uniform strength to the $w$-plane does not yield these features; it actually creates a velocity profile in the nozzle channel which achieves its maximum velocity on the channel edges and its minimum velocity in the channel center.  We rectify this by making the substitution $m \rightarrow m\sin{(\arg{z})}$.  This yields the velocity profile shown in black points in Fig.~\ref{channel_profile}.  The blue line represents a parabolic fit to the velocity profile; we see that this updated model has a velocity profile that is very nearly quadratic in nature, reflecting what previous experimental work has shown to be true.

Bringing this all together, we have a theoretical prediction for flow through the nozzle and into the reservoir.  The model depends on the free parameter $m$, which represents the speed at which smectic material is flowing through the nozzle.

\begin{figure}
\centering
\includegraphics[keepaspectratio=true,width=0.5\textwidth]{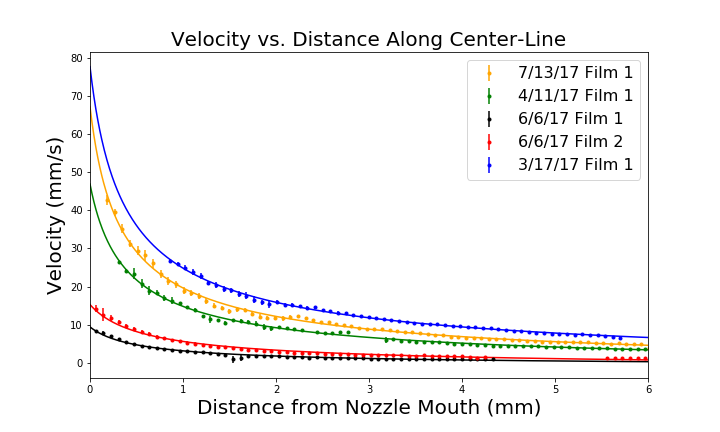}
\caption{Velocity vs. distance along the the center line. The velocity being plotted is the magnitude of the velocity. The origin is located at the nozzle mouth. The velocity falls off as a function of $dr=y^{-1/3}$ .}
\label{par}
\end{figure}

\begin{figure}
\centering
\includegraphics[keepaspectratio=true,width=0.5\textwidth]{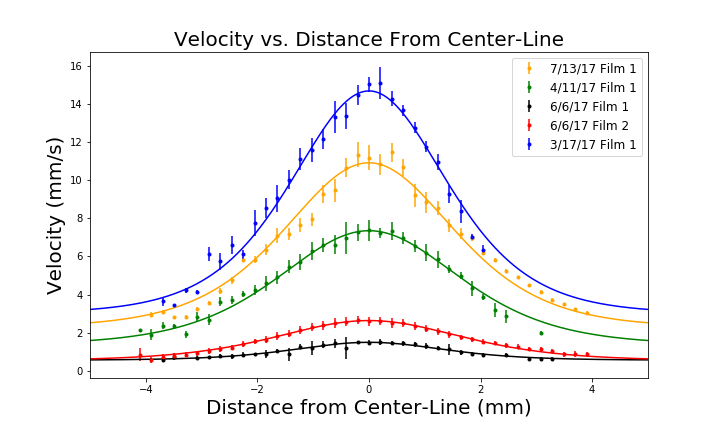}
\caption{Velocity vs. distance perpendicular to the center line at 4.3 mm from the nozzle mouth. Again, the velocity bieng ploted is the magnitude of the velocity. The velocity profile displays the same behavior for a wide range of initial velocities. The velocity is given by $dr= asech^{2}{(b*x)}$}
\label{results}
\end{figure}

\section{Results and Discussion}

To  determine  the  validity  of  our  model,  we  compare experimental  data  to  theoretical  predictions  along  two characteristic lines:  one extending out from the center of the nozzle into the reservoir (referred to as the center-line) and one perpendicular to this line (taken to be 4.3 mm from the nozzle mouth). As the distance from the nozzle mouth increases, the maximum particle velocity decreases exponentially, as seen in  Fig.~\ref{par}. Perpendicular to the nozzle mouth, we see that the velocity is given by a $sech^{2}$ distribution, as seen in Fig.~\ref{results}.  The results apear to agree with the theory, indicating that smectic A films provide an ideal testbed for 2D laminar hydrodynamics. 
\begin{figure}
\centering
\includegraphics[keepaspectratio=true,width=0.47\textwidth]{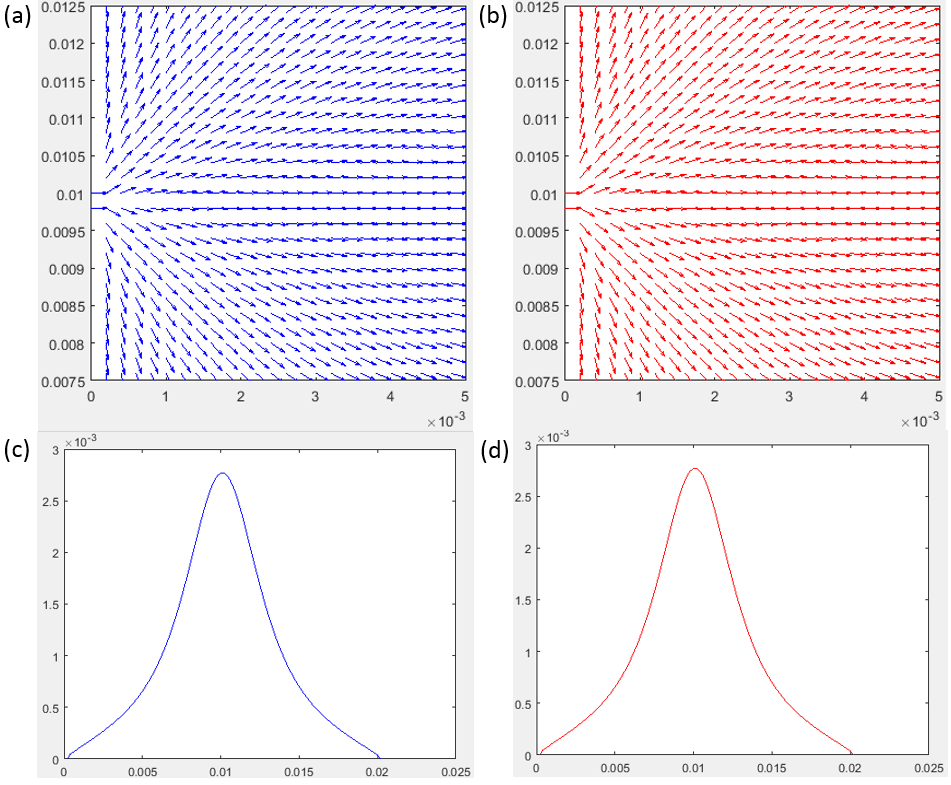}
\caption{Simulation results of velocity profile near the two dimensional laminar jet with (a) and without (b) air. Velocity vs. distance perpendicular to the center line at 4 mm from the nozzle mouth with (c) and without air(d).}
\label{WithWithouAir}
\end{figure}
We see that the model is able to accurately predict velocities along the center-line; however, there are some small issues along the perpendicular line.  Most noticeable is the fact that the data is slightly shifted to the left.  This is because the meniscus at the mouth of the reservoir is slightly larger on the right side than on the left, shifting the flow ever so slightly to the left.  Even if the data were centered though, the model still fails to predict quite as fast of a fall-off in velocity with distance as is observed.  Presumably, an expression for the strength of the pre-transformation source that is more complex than the ``$m\sin{(\arg{z})}$" would yield more accurate results.

Since our 2D laminar jet flow experiments were carried out in the air, it is natural for readers to question whether our models of neglecting the air-film frictional force is over simplifying or not. Our previous experiments~\cite{qi2016experimental} showed the air friction placed a significant role in confining the movement of inclusion within the FSLCFs, one may ask whether this air-film friction will also play a similar role in determined the flow field or the shape of the streamline? Computer simulations that built upon Seibold~\cite{seibold_2008} were carried out with the assumption that air friction is solely from the shearing force between the moving film and the air, with the boundary condition that the air movement at the container is zero. Also, the flow speed at the nozzle is set to be 0.1 m/s. With pressure correction approach to solve the 2D incompressible Navier-Stokes Equations numerically, we obtained the velocity profiles near the nozzle with (Fig.~\ref{WithWithouAir}(a)) and without air(Fig.~\ref{WithWithouAir}(b)). We also  study velocity
vs. distance perpendicular to the center line at 4 mm from
the nozzle mouth with (Fig.~\ref{WithWithouAir}(c)) and without air(Fig.~\ref{WithWithouAir}(d)). The simulations indicate air friction has almost negligible effect on the velocity maps of flow near the nozzle.

This seemingly stunning results could actually be understood in analogy with the case of 3D incompressible Newtonian fluids. In a water pipe, for example, the inner pressure has little effect on the flow rate, since it is the pressure difference rather than the absolute pressure determine the flow profile (Poiseuille's law). In our case, due to the 2D incompressibility of the smectic films, the external air frictional force could be quickly dissipated and equalized through every point, thus make the flow profile unchanged.

\section{Conclusion}
Despite small inaccuracies in the model's predictions of velocities along the perpendicular line, the model is overall successful in describing the behavior of this system.  Because the model assumes a completely inviscid fluid, this result lends itself to the perhaps surprising conclusion that, when it comes to long-range flows, MX~$12805$ films behave as if they have negligible viscosity.  Future testing will probe this effect on other smectic materials and at varying speeds, attempting to determine the regime in which even a low viscosity becomes important to the dynamics at play.

This work was supported by NASA Grant~NNX-13AQ81G and by the Soft Materials Research Center under NSF MRSEC Grants~DMR-0820579 and DMR-1420736.

\bibliographystyle{apsrev}
\bibliography{references} 

\end{document}